\documentclass[10pt,reqno]{article}

\usepackage{ulem}

\usepackage{amsmath,amssymb,amsthm}
\usepackage{esint}
\usepackage{bm} 
\usepackage{url}
\usepackage[sort, numbers]{natbib}
\usepackage{subfigure}
\usepackage{graphicx,color}
\usepackage{algorithm}
\usepackage{algpseudocode}
\usepackage{algorithmicx}
\usepackage{hyperref}
\usepackage{float}
\usepackage{booktabs,multirow}
\usepackage{hhline}
\usepackage{setspace}

\usepackage{caption}
\usepackage{hyphsubst}
\usepackage{caption}

\usepackage{float}
\usepackage{hyperref}

\hypersetup{colorlinks,citecolor=blue,linkcolor=blue, urlcolor=blue}
\usepackage{pdflscape}
\def\Bj{{\bf j}}

\def\Bn{{\bf n}}

\def\Bp{{\bf p}}

\def\Bx{{\bf x}}
\def\By{{\bf y}}

\newcommand{\RR}{\mathbb{R}}
\newcommand{\ds}{\displaystyle}
\def\bx{{{\bf x}}}

\def\hbx{{\hat{\bx}}}

\def\by{{{\bf y}}}
\def\bel{{\boldsymbol{\ell}}}

\def\CJ{{\mathcal{J}}}
\def\CE{{\mathcal{E}}}
\def\CH{{\mathcal{H}}}
\def\BJ{{\bf J}}
\def\BE{{\bf E}}
\def\BH{{\bf H}}
\def\BF{{\bf F}}
\DeclareMathAlphabet{\itbf}{OML}{cmm}{b}{it}

\theoremstyle{plain}
\newtheorem{theorem}{Theorem}[section]

\theoremstyle{definition}
\newtheorem{definition}[theorem]{Definition}
\newtheorem{problem}[theorem]{Problem}

\theoremstyle{remark}

\newcommand{\email}[1]{\protect\href{mailto:#1}{#1}}

\newcommand{\pathfigures}{Figures/}
\graphicspath{{\pathfigures}}

\date{} 
\begin{document}
	
\title{Multi-Frequency Far-Field Data Enrichment for Electromagnetic Source Reconstruction %\footnotemark[1]
}
	\author{
		Atyab Khalifa Al-Shaqsi\footnotemark[1]
		\and
        Heba Mohammed Al-Subhi\footnotemark[1]
		\and
        Xianchao Wang\footnotemark[2]
        \and
        Shujaat Khan\footnotemark[3]
		\and
		Abdul Wahab\footnotemark[1]\, \footnotemark[4] 
	}
	\maketitle
	
	\renewcommand{\thefootnote}{\fnsymbol{footnote}}
	%\footnotetext[1]{The work was supported by the Grant.}
    \footnotetext[1]{Department of Mathematics, College of Science, Sultan Qaboos University, Muscat 123, Oman (\email{s135332@student.squ.edu.om}; \email{s135318@student.squ.edu.om}; \email{a.wahab@squ.edu.om}).}
    \footnotetext[2]{School of Mathematics, Harbin Institute of Technology, Harbin, P. R. China (\email{xcwang90@gmail.com};  \email{xcwang@hit.edu.cn}).}
    \footnotetext[3]{Department of Computer Engineering, College of Computing and Mathematics, King Fahd University of Petroleum \& Minerals, Dhahran 31261, Saudi Arabia (\email{shujaat.khan@kfupm.edu.sa}).}
	\footnotetext[4]{Corresponding author.}%: A. Wahab (\email{a.wahab@squ.edu.om}).
	\renewcommand{\thefootnote}{\arabic{footnote}}
\begin{abstract} 
		Reconstructing unknown electromagnetic sources from far-field radiation patterns is a fundamental inverse problem with broad applications in biomedical imaging, non-destructive testing, and telecommunications. In practical settings, however, collecting dense multi-frequency far-field measurements at the Nyquist sampling rate is often infeasible. Under-sampled or sparse data introduce non-radiating source components that sever the uniqueness of the solution, creating severe artifacts when standard inversion techniques are applied. To overcome this limitation, we present a two-stage reconstruction strategy exploiting the physical property that compactly supported, geometrically sparse sources exhibit a finite rate of innovations (FRI). In the first stage, we construct an associated wrap-around structured Hankel matrix. By leveraging the low-rank property of the matrix due to FRI of the unknown sources, we enrich the sub-sampled data. To that end, we convert missing multi-frequency far-field data recovery into a constrained matrix completion task solved via Annihilating Filter-based Low-rank Hankel Matrix Completion Approach (ALOHA). In the second stage, a Fourier inversion scheme reconstructs the current source density from the enriched dataset. Extensive numerical evaluations on electromagnetic source models show that our enrichment framework effectively eliminates under-sampling artifacts and resolves non-uniqueness challenges. The method delivers accurate and stable reconstructions under high sub-sampling rates (e.g., with $30$\% to $50$\% available samples) and strong noise conditions ($10$ dB SNR), outperforming standard $\ell_1$-compressed sensing baselines.
	\end{abstract}
	
	
\section{Introduction}
	
Reconstructing unknown electromagnetic sources from far-field measurements is a cornerstone problem in applied physics and computational engineering \cite{ISPs}. The ability to infer internal current distributions from external radiation patterns underpins a wide range of technologies. In biomedical imaging, for instance, inverse source models enable electroencephalography (EEG) and magnetoencephalography (MEG) to map functional brain activities such as neuronal responses for intracranial recordings and localization of cortical activity \cite{Thio23, Beltrachini21, Jerbi04, Ammari02, multipolar}. Also, wave-scattering principles allow clinicians to locate and characterize tumors non-invasively \cite{Ammari-multi}. Beyond medical diagnostics, inverse source problems (ISPs) are equally vital in non-destructive testing for structural health monitoring \cite{Takahashi22}, security robotics \cite{AmmariN}, and in conformal antenna design for synthesizing complex radiation fields \cite{Leone18}.

Over the years, a variety of theoretical and numerical frameworks have been developed to address electromagnetic ISPs based on their possible uses, experimental configurations, practical requirements or source kinds.
Devaney, Marengo, and Li \cite{Devaney} mathematically studied ISPs and formulated an optimization strategy to reconstruct minimal-energy source distribution in inhomogeneous media. Eibert, Vojvodi\'c, and Hansen \cite{Eibert16} introduced an approach based on Huygens radiator to identify directional Gaussian-beam-like sources. 
Griesmaier and Schmiedecke \cite{Griesmaier017b} adopted a \textit{Multiple Signal Classification} (MUSIC) algorithm for ISPs. In a complete Maxwell equations context, a time-reversal algorithm was presented in \cite{Wahab14}. A stochastic model to reconstruct random or fluctuating current sources has been studied in \cite{Bao14}. Recently, Wang et al. \cite{WangEM}
proposed an efficient Fourier inversion algorithm capable of accurate source reconstruction from dense multi-frequency far-field data. 

Despite these theoretical advances, practical applications face a major hurdle that physical constraints, restricted measurement apertures, and sensor costs frequently make it impossible to collect continuous or dense multi-frequency data at the Nyquist-sampled grid. When measurement grids are sparse or under-sampled, ISPs become severely ill-posed. Under-sampling masks specific frequency components, causing parts of the source to behave like non-radiating modes whose radiation signatures never reach the detectors. As the non-radiating modes lie within the null space of the forward wave operator, conventional inversion techniques suffer from severe non-uniqueness, generating heavy ringing and speckle artifacts \cite{Bleistein, Hanke1, Hanke2}. 
Traditional workarounds, such as least-squares-based minimum energy solutions, Tikhonov or Total Variation (TV) regularization techniques or standard $\ell_1$-compressed sensing techniques \cite{Saman, Eller09, Bao15IP, Bao15SNA, Griesmaier17} often fail to resolve this issue, particularly when under-sampling artifacts match or exceed the scale of the source features themselves. To bridge this gap, we present a data enrichment strategy in this work designed specifically for sparse multi-frequency far-field datasets. 

Instead of applying ad-hoc regularization directly to an ill-conditioned inverse operator, we focus on enriching the under-sampled data by relating it to the Fourier spectrum of the unknown source. Assuming sparse grid under-sampled dataset being a random subset of a Nyquist-sampled grid dataset, we frame an optimization problem for the recovery of the \textit{missing data} (i.e., the data enrichment) problem. Recognizing that many physical electromagnetic sources are spatially compact and geometrically sparse in the physical domain, we reframe the data enrichment problem through the lens of Finite Rate of Innovations (FRI) signal processing \cite{vetterli2002sampling}. Using Helmholtz-Hodge decomposition \cite{Bhatia, Lindell}, we decouple the unknown source field into transverse electric and transverse magnetic spectral components that are directly proportional to the measurements. The transverse spectral components are mapped into a block-diagonal, wrap-around Hankel matrix whose low-rank property stems from the FRI property of the transverse components. This  allows us to formulate missing data recovery as a constrained rank-minimization problem, which we solve using \textit{Annihilating Filter-based Low-rank Hankel Matrix Completion Approach (ALOHA)} via the \textit{Alternating Direction Method of Multipliers (ADMM)} \cite{ye2016compressive, jin2015annihilating, jin2016general}. The enriched dataset is then passed on to the Fourier inversion scheme to reconstruct the current source density cleanly. The choice of ALOHA among other matrix completion approaches is purely based on its stability and robustness. Extensive numerical evaluations have been done on variety of current source models, sampling rates and measurement noise conditions to substantiate the superiority of the proposed strategy over the standard $\ell_1$-compressed sensing baseline algorithm.

The rest of this article is organized as follows: In Section \ref{sec:Form}, we provide a detailed mathematical formulation of the electromagnetic ISP with sparse multi-frequency data. Section \ref{sec:Research} provides the details of the data enrichment strategy. First the Fourier algorithm for dense datasets is presented in Section \ref{ss:Fourier}. Then the idea of the data enrichment framework is provided in Section \ref{ss:Enrichment}, and finally the constraint optimization problem for data enrichment is formulated in Section \ref{ss:Optimization}. In Section \ref{sec:Setup}, experimental setup and evaluation protocols are elaborated. The quantitative and qualitative analysis is done in \ref{sec:Num}. The article concludes with Section \ref{sec:Conc} containing the summary of the results.
	
\section{Mathematical formulation}\label{sec:Form} 

This article is concerned with an ISP related to electromagnetic wave propagation in the free domain. To facilitate ensuing discussion, we first discuss some preliminaries.  

\subsection{Governing Equations} 

The electromagnetic wave propagation in homogeneous isotropic free domain $\RR^3$, characterized by electric permittivity $\varepsilon>0$  [F$\cdot$m$^{-1}$] and magnetic permeability $\mu>0$ [H$\cdot$m$^{-1}$], respectively, due to an electric current source density $\CJ:\RR^3\times \RR\to \RR^3$ [Am$\cdot$ m$ ^{-2}$], is governed by the Maxwell equations, 
\begin{align}
    \begin{cases}
        -\varepsilon \dfrac{\partial \CE}{\partial t} (\bx,t)+\nabla \times \CH (\bx,t) = \CJ(\bx,t), &\bx\in\RR^3, \,\, t>0, 
        \\[0.8em]
        \mu\dfrac{\partial \CH}{\partial t}(\bx,t)+\nabla \times \CE(\bx,t) =\mathbf{0}, &\bx\in\RR^3, \,\, t>0.
    \end{cases}\label{Mt}
\end{align}
Here, $\CE:\RR^3\times \RR\to\RR^3$ and $\CH:\RR^3\times \RR\to\RR^3$ are the electric field [Volt$\cdot$m$^{-1}$] and magnetic field [Am$\cdot$m$^{-1}$] in $\RR^3$, respectively.  
 With appropriate regularity and suitable initial conditions,  Maxwell equations \eqref{Mt} possesses a unique solution $(\CE,\CH)$. A detailed study of the existence of a unique solution to \eqref{Mt} can be found in the monographs \cite{Nedelec, Monk}.

Using Fourier transform, the time-dependent Maxwell equations \eqref{Mt} can be converted to time-harmonic equations, 
\begin{align}
    \begin{cases}
        i\omega\varepsilon \BE(\bx,\omega)+\nabla \times \BH (\bx,\omega) = \Bj(\bx,\omega), &\bx\in\RR^3, 
        \\[0.8em]
        -i\omega\mu\BH(\bx,\omega)+\nabla \times \BE(\bx,\omega) =\mathbf{0}, &\bx\in\RR^3,
    \end{cases}\label{Momega}
\end{align}
where $\omega>0$ is the radial frequency [Hz]  and 
$$
    \BE(\bx,\omega)=\int_\RR\CE(\bx, t)e^{-i\omega t}dt, \quad 
    \BH(\bx,\omega)= \int_\RR\CH(\bx, t)e^{-i\omega t}dt, \quad
    \BJ(\bx,\omega):= \int_\RR \CJ(\bx,t)e^{-i\omega t}dt.
$$
For well posedness of the time-harmonic system, it is customary to impose Silver-M\"{u}ler radiation conditions, $\hat{\bx}:=\dfrac{\bx}{|\bx|}$ and $\bx\neq 0$,
\begin{align}
    \lim_{|\bx|\to\infty} 
    \begin{cases}
    \sqrt{\mu} \BH(\hat{\bx},\omega) \times \hat{\bx} -\sqrt{\varepsilon} \BE(\hat{\bx},\omega)=\mathbf{0},
    \\[0.8em]
     \sqrt{\varepsilon} \BE(\hat{\bx},\omega) \times \hat{\bx} +\sqrt{\mu} \BH(\hat{\bx},\omega)=\mathbf{0}.
     \end{cases}\label{RC}
\end{align}

Throughout this article, we consider stationary (in time) compactly supported (in space) current source density, i.e., $\CJ(\bx,t)=\BF(\bx)\delta_0(t)$ for some smooth function $\BF:\RR^3\to\RR^3$ such that there exists a compact domain $D\subset \RR^3$ such that 
$$
    \text{supp}\{\BF\}:= \overline{\Big\{\bx\in\RR^3\,\,: \, \BF(\bx)\neq \mathbf{0}\Big\}}\subset D\subset \RR^3.
$$
Here, $\delta_0$ is the \textit{Dirac mass} at $t=0$ and the superposed bar indicates the closure. Owing to the definition of $\CJ(\bx,t)$, the time-harmonic current source density $\BJ$ is given by $\BJ(\bx,\omega)=\BF(\bx)$.

We can uncouple the Maxwell equations \eqref{Momega} to get the electric and magnetic formulations of the Maxwell equations, 
$$
    \begin{cases}
        \nabla \times \nabla \times \BE (\bx,\omega) - \kappa^2\BE (\bx,\omega) = i\omega \mu \BF(\bx), &\bx\in\RR^3,
    \\
        \nabla \times \nabla \times \BH(\bx,\omega) - \kappa^2\BH(\bx,\omega)=\nabla \times \BF(\bx), & \bx\in\RR^3.
    \end{cases}
$$
The parameter $\kappa=\omega/c>0$ is the electromagnetic \textit{wavenumber}, where we define the \textit{speed of light} [m$\cdot$s$^{-1}$] in the electromagnetic medium by $c:={1}/{\sqrt{\mu\varepsilon}}$. 

\subsection{Far-Field Expansions} 

The Silver-M\"{u}ler radiation conditions \eqref{RC} imposed on the electric and magnetic field force them to have a smooth behavior far away from the source $\BF$ \cite{Nedelec, Monk, Colton-Kress}. Specifically, it ensures the existence of two analytic functions $\BE_\infty:\mathbb{S}^2\to \RR^3$ and $\BH_\infty:\mathbb{S}^2\to \RR^3$ such that 
$$
    \BE(\bx,\omega)=\dfrac{e^{i\kappa|\bx|}}{|\bx|}\left\{\BE_\infty(\hat{\bx},\omega)+O\left(\dfrac{1}{|\bx|}\right)\right\}
    \,\text{ and }\,
     \BH(\bx,\omega)=\dfrac{e^{i\kappa|\bx|}}{|\bx|}\left\{\BH_\infty(\hat{\bx},\omega)+O\left(\dfrac{1}{|\bx|}\right)\right\},
$$
as $|\bx|\to +\infty$, where $\mathbb{S}^2:=\left\{\bx\in\RR^3:\,|\bx|=1\right\}$ and a superposed hat represents a unit vector. 
The functions $\BE_\infty$ and $\BH_\infty$ are called the electric and magnetic far-fields as they describe the far-field signatures of the electric and magnetic fields produced by $\BF$. Note that $\hbx \cdot \BE_\infty(\hbx) = 0 = \hbx\cdot \BH_\infty(\hbx)$. It is important to mention that some current sources may be so weak that their far-field signature $(\BE_\infty, \BH_\infty)$ vanishes. We refer to the monograph by Colton and Kress \cite{Colton-Kress} for more details on the topic. 

\subsection{Helmholtz-Hodge Decomposition of the Current Density}

\begin{figure}[!htb]
    \centering
    \includegraphics[width=0.45\linewidth]{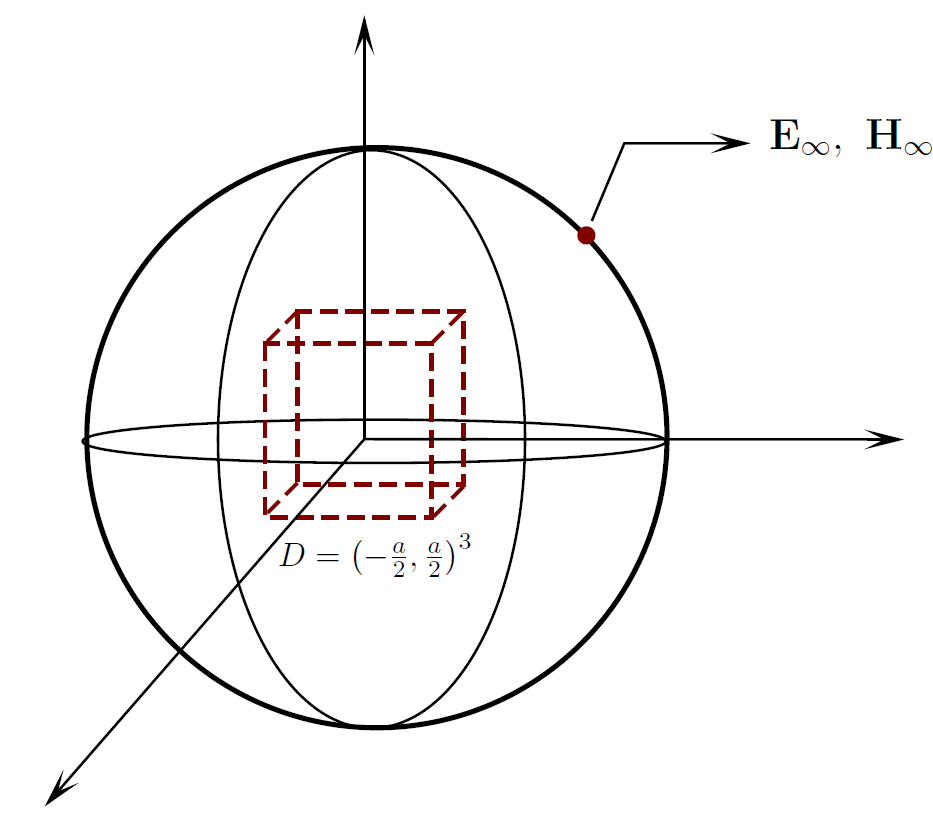}
    \caption{Illustration of the geometry.}
    \label{geometry}
\end{figure}

As specified earlier, we assume that the current density $\BF$ is compactly supported in the bounded domain $D$. In the rest of this work, we take $D$ to be a cube with side length $a>0$ [m], i.e.,
$$
    D:=\left(-\dfrac{a}{2},\dfrac{a}{2}\right)^3, \qquad a\in\RR,\,\, a>0.
$$
An illustration of the unit sphere $\mathbb{S}^2$ and the domain $D$ is provided in Fig. \ref{geometry}

Let $\hat{\Bp}$ be the direction of polarization. Then $\BF$, under certain smoothness conditions, can be resolved as the sum of the components along and orthogonal to the direction $\hat{\Bp}$. Specifically, there exist two scalar fields, $f:\RR^3 \to \RR$ and $g:\RR^3\to \RR$, for a direction $\hat{\Bp}$ such that $\BF$ can be expressed as
\begin{equation}
    \BF(\bx) = \hat{\Bp} f(\bx)+\hat{\Bp} \times \nabla g(\bx). \label{Helm-Decomp}
\end{equation}
We call $f$ and $g$ the transverse electric and magnetic components of $\BF$, respectively. Note that 
$$
    \BF(\bx)\cdot \hat{\Bp}= f(\bx)\quad\text{and}\quad \BF(\bx)\times \hat{\Bp} = (\hat{\Bp}\cdot \hat{\Bp})\nabla g(\bx)-(\hat{\Bp}\cdot \nabla g(\bx))\hat{\Bp} =\nabla g(\bx).
$$
We refer to \cite{Lindell, Bhatia} for more details on this decomposition. 

Throughout this article, the polarization direction $\hat{\Bp}$ is assumed to be known and is taken from the set of admissible polarization directions 
$$
    \mathbb{P}:=\Big\{\hat{\Bp}\in\mathbb{S}^2:\, \hat{\Bp}\times \bel\neq \textbf{0}, \,\, \bel\in\mathbb{Z}^3\setminus\{\mathbf{0}\}\Big\},
$$
where $\mathbb{Z}$ is the set of integers. 

\subsection{Inverse Problem}
To present the ISP at hand, we first give some notation. 
\begin{definition}[Admissible data points]
Let $\epsilon_0\in\RR_+$ be such that $\epsilon_0\to 0^+$. We introduce the multi-index $\bel_0:=(\epsilon_0,0,0)$. We define the  \textit{admissible data points} 
\begin{align}
\hat{\bx}_{\bel}:=&
\begin{cases}
\hat{\mathbf{\bel}}=\dfrac{\bel}{|\bel|}, &\mathbf{\bel}\in\mathbb{Z}^3\setminus\{\mathbf{0}\},
\\[1em] 
\hat{\mathbf{\bel}}_0=\dfrac{\bel_0}{|\bel_0|}=(1,0,0), & \mathbf{\bel}=\mathbf{0}, 
\end{cases}\label{xl}
\end{align}
and the \textit{admissible wavenumbers} by
\begin{align}
\kappa_{\mathbf{\bel}}:=& \dfrac{2\pi}{a}
\begin{cases}
|\bel|, &\mathbf{\bel}\in\mathbb{Z}^3\setminus\{\mathbf{0}\},
\\[0.5em]
\epsilon_0, & \mathbf{\bel}=\mathbf{0}.
\end{cases}\label{kl}
\end{align}
We also define the \textit{admissible wavelengths} as  $\lambda_\bel:=1/\kappa_\bel$. 
\end{definition}

\begin{definition}[Dense and sparse sampling points]
Let us define sampling grid on the unit sphere as 
$$
\mathbb{S}^{2}_{\rm dis}:=\left\{\hat{\bx}_{\mathbf{\bel}}\in\mathbb{S}^{2}\, \left|\right.\quad \mathbf{\bel}\in\mathbb{Z}^3\right\}.
$$

Let $\Omega_M:=\{\hat{\bx}_1,\cdots,\hat{\bx}_M\}\subset \mathbb{S}^{2}_{\rm dis}$
be a \textit{dense set} of Nyquist-sampled grid points, i.e., the sampling points, $\hat{\bx}_m$  defined by \eqref{xl}, for $m=1,\cdots, M$, are chosen at most $\lambda_{\rm min}/2$ distant apart from each other, where $\lambda_{\rm min}:= \min\{\lambda_1,\cdots, \lambda_M\}$.
We also define the \textit{sparse set} of measurement points, $\Omega_{M_R}$, a randomly chosen subset of $\Omega_{M}$ for $R\ll M$, i.e.,   
$$
\Omega_{M_R}:=\{\hat{\bx}_{M_r}\,|\quad r=1,\cdots, R\}\subset\Omega_M\subset \mathbb{S}^{2}_{\rm dis}.
$$
\end{definition}

In this work, we discuss the following ISPs. 

\begin{problem}[Inverse problem with full measurements]\label{probF}
Given a polarization direction $\hat{\Bp}\in\mathbb{P}$, find the current source density $\BF(\bx)$ for the Maxwell equations \eqref{Momega} subject to Silver-M\"{u}ler radiation conditions \eqref{RC} given the multi-frequency far-field data 
  $$
  \Big\{\BE_\infty(\hat{\bx}_{m}, \kappa_{m})\,\Big| \,\,\hat{\bx}_{m}\in\Omega_{M},\,\, m=1,\cdots,M\Big\},
  $$
  or
  $$
  \Big\{\BH_\infty(\hat{\bx}_{m}, \kappa_{m})\,\Big| \,\,\hat{\bx}_{m}\in\Omega_{M},\,\, m=1,\cdots,M\Big\}.
  $$
\end{problem}

\begin{problem}[Inverse problem with sparse measurements]\label{probS}
Given a polarization direction $\hat{\Bp}\in\mathbb{P}$, find the current  source density $\BF(\bx)$ for the Maxwell equations \eqref{Momega} subject to Silver-M\"{u}ler radiation conditions \eqref{RC} given the multi-frequency far-field data 
  $$
  \Big\{\BE_\infty(\hat{\bx}_{M_r}, \kappa_{M_r})\,\Big|\quad \hat{\bx}_{M_r}\in\Omega_{M_R}, \,\, r=1,\cdots, R\Big\},
  $$
  or 
  $$
  \Big\{\BH_\infty(\hat{\bx}_{M_r}, \kappa_{M_r})\Big)\,\Big|\,\, \hat{\bx}_{M_r}\in\Omega_{M_R}, \,\, r=1,\cdots, R\Big\}.
  $$
\end{problem}

\section{Matrix Completion Approach for Data Enrichment}\label{sec:Research}

For solving Problem \ref{probS}, we aim to enrich the sparse data before a Fourier inversion. The core of our data enrichment strategy relies on modeling the missing spectral components of the unknown current source through the framework of FRI signal processing \cite{vetterli2002sampling}. By mapping the available multi-frequency data into a structured, block-diagonal Hankel matrix, we recast spectral completion as a constrained matrix completion problem under a low-rank constraint. We then solve this optimization problem using the ALOHA \cite{ye2016compressive}. The details are provided below.

\subsection{Fourier Algorithm for Full Measurement Problem}\label{ss:Fourier}

When the dense far-field dataset is available, the solution $\BF$ of the Problem \ref{probF} is given as 
\begin{equation}
\label{Fourier-F}
    \BF(\bx)=\sum_{\bel\in\mathbb{Z}^3}\hat{\Bp}{f}_\bel \varphi_{\bel}+\dfrac{2\pi i}{a}\sum_{\bel\in\mathbb{Z}^3\setminus\{\mathbf{0}\}}(\hat{\Bp}\times \bel) {g}_\bel\varphi_{\bel},
\end{equation}
by a Fourier inversion. Here
$$
    \begin{cases}
    f(\bx):=\ds\sum_{\bel\in\mathbb{Z}^3} {f}_\bel \varphi_{\bel}(\bx), & \ds{f}_\bel:=\dfrac{1}{a^3}\int_{D}f(\bx)\overline{\varphi_\bel(\bx)}d\bx,
    \\
    g(\bx):=\ds\sum_{\bel\in\mathbb{Z}^3\setminus\{\mathbf{0}\}} {g}_\bel\varphi_{\bel}, &{g}_\bel:=\ds\dfrac{1}{a^3}\int_{D}g(\bx)\overline{\varphi_\bel(\bx)}d\bx,
    \\
    \varphi_{\bel}(\bx):=\ds\exp\left(\dfrac{2\pi i}{a}\hat{\bx}_{\bel}^T\bx\right), &\bel\in\mathbb{Z}^3, \,\,\bx\in\RR^3,
    \end{cases}
$$
where superposed bar represents complex conjugate and $f_\bel$ and $g_\bel$ in \eqref{Fourier-F} are the Fourier coefficients of the transverse electric and magnetic components $f$ and $g$ of the current source density $\BF$ as in \eqref{Helm-Decomp}.

Fourier coefficients $f_\bel$ and $g_\bel$ can be obtained by using the electric or magnetic far-field data. In fact,  the coefficients $f_\bel$ and $g_\bel$ are related to the magnetic far-field measurements as 
\begin{align}
    f_\bel&= \dfrac{4\pi\hat{\bx}_\bel\times\hat{\Bp}\cdot\BH_\infty(\hat{\bx}_\bel,\kappa_\bel)}{i\kappa_\bel a^3 |\hat{\bx}_\bel \times \hat{\Bp}|^2} \quad\text{and}\quad
    g_\bel=- \dfrac{2\hat{\bx}_\bel\times(\hat{\Bp}\times \bel)\cdot\BH_\infty(\hat{\bx}_\bel,\kappa_\bel)}{\kappa_\bel a^2 |\hat{\bx}_\bel \times (\hat{\Bp}\times \bel)|^2},\label{flh}
    \\
    f_{0} &= \dfrac{\epsilon_0\pi}{a^3\sin (\epsilon_0 \pi)}\left(\dfrac{4\pi\hbx_{\bel_0}\times\hat{\Bp}\cdot\BH_\infty(\hbx_{\bel_0},\kappa_{\bel_0})}{i\kappa_{\bel_0}|\hbx_{\bel_0}\times\hat{\Bp}|^2}-\sum_{\bel\in\mathbb{Z}^3\setminus\{\mathbf{0}\}} f_\bel \int_D\exp\left(i(\kappa_\bel\hat{\bx}_{\bel}^T-\kappa_{\bel_0}\hbx_{\bel_0}^T)\by\right)d\by\right), \label{flh0}
\end{align}
and to the electric far-field measurements as 
\begin{align}
    f_\bel&= \dfrac{4\pi\hat{\Bp}\cdot\BE_\infty(\hat{\bx}_\bel,\kappa_\bel)}{i\omega\mu a^3 |\hat{\bx}_\bel \times \hat{\Bp}|^2} \quad\text{and}\quad
    g_\bel=- \dfrac{2(\hat{\Bp}\times \bel)\cdot\BE_\infty(\hat{\bx}_\bel,\kappa_\bel)}{\omega \mu a^2 |\hat{\bx}_\bel \times (\hat{\Bp}\times \bel)|^2},\label{fle}
\\
    f_{0} &= \dfrac{\epsilon_0\pi}{a^3\sin (\epsilon_0 \pi)}\left(\dfrac{4\pi\hat{\Bp}\cdot\BE_\infty(\hbx_{\bel_0},\kappa_{\bel_0})}{i\omega \mu|\hbx_{\bel_0}\times\hat{\Bp}|^2}-\sum_{\bel\in\mathbb{Z}^3\setminus\{\mathbf{0}\}} f_\bel \int_D\exp\left(i(\kappa_\bel\hat{\bx}_{\bel}^T-\kappa_{\bel_0}\hbx_{\bel_0}^T)\by\right)d\by\right),\label{fle0}
\end{align}
depending on the measurements at hand \cite{WangEM}. 
This shows that the current source density can be computed from the discrete far-field signatures $\BE_\infty(\hbx_\bel,\kappa_\bel)$ or $\BH_\infty(\hbx_\bel,\kappa_\bel)$. However, to solve the Problem \ref{probF}, we have to truncate the Fourier series up to order $N\in\mathbb{N}$ so that $M=(2N+1)^3$ and can compute only the coefficients $f_{\bel}$ and $g_\bel$ such that $1\leq \|\bel\|_\infty \leq N$, where $\|\cdot\|_\infty$ is the infinity norm. Thus, an approximate solution can be obtained from the dense dataset by 
$$
    \BF_{M}\approx\hat{\Bp} f_0+\sum_{1\leq \|\bel\|_\infty \leq N}\left(\hat{\Bp} f_{\bel}+\dfrac{2\pi i}{a}(\hat{\Bp}\times \bel)g_{\bel}\right)\varphi_{\bel},\label{FM}
$$
where $f_0$  is approximated as
$$
    f_{0} \approx \dfrac{\epsilon_0\pi}{a^3\sin (\epsilon_0\pi)}\left(\dfrac{4\pi\hbx_{\bel_0}\times\hat{\Bp}\cdot\BH_\infty(\hbx_{\bel_0},\kappa_{\bel_0})}{i\kappa_{\bel_0}|\hbx_{\bel_0}\times\hat{\Bp}|^2}-\sum_{1\leq \|\bel\|_\infty \leq N} f_\bel \int_D\exp\left(i(\kappa_\bel\hat{\bx}_{\bel}^T-\kappa_{\bel_0}\hbx_{\bel_0}^T)\by\right)d\by\right),%\label{f0lh}
$$
or 
$$
    f_{0} \approx \dfrac{\epsilon_0\pi}{a^3\sin (\epsilon_0\pi)}\left(\dfrac{4\pi\hat{\Bp}\cdot\BE_\infty(\hbx_{\bel_0},\kappa_{\bel_0})}{i\omega \mu|\hbx_{\bel_0}\times\hat{\Bp}|^2}-\sum_{1\leq \|\bel\|_\infty \leq N} f_\bel \int_D\exp\left(i(\kappa_\bel\hat{\bx}_{\bel}^T-\kappa_{\bel_0}\hbx_{\bel_0}^T)\by\right)d\by\right).
$$

\subsection{Data Enrichment Strategy}\label{ss:Enrichment}

When dense multi-frequency far-fields are recorded across a Nyquist-sampled grid $\Omega_M$, a \textit{sufficiently} large number of Fourier coefficients $f_\bel$ and $g_\bel$ can be determined directly using Eqs. \eqref{flh}--\eqref{fle0} rendering an accurate approximation $\BF_ M$ of the true current source density $\BF$. However, in practical situations, complete Nyquist-rate sampling is rarely feasible due to physical constraints, sensor costs, and restricted apertures. For sub-sampled measurement grid $\Omega_{M_R} \subset \Omega_M$, incomplete data, particularly inaccuracies in estimating the zero-frequency mode $f_0$, significantly reduce reconstruction fidelity and add severe visual artifacts. Practically, unrecorded far-field data at \textit{missing} sampling points, $\hbx_m \in \Omega_M \setminus \Omega_{M_R}$, leave certain spatial frequencies unaccounted for. Consequently, parts of the source behave as non-radiating modes at these unobserved frequencies. Since non-radiating components reside in the null space of the forward wave operator, conventional direct inversion techniques suffer from severe non-uniqueness, manifested as heavy speckle and ringing artifacts  \cite{jin2015annihilating}. Traditional remedies relying on standard Tikhonov, Total Variation (TV), or $\ell_1$-compressed sensing regularization techniques \cite{Fannjiang} often struggle in these scenarios, particularly when the spatial extent of the under-sampling artifacts matches or exceeds that of the source features themselves.

To overcome these challenges, we leverage the spatial compactness and geometric sparsity of $\BF$ inside $D$ to recover the unobserved far-field data at $\hbx_m \in \Omega_M \setminus \Omega_{M_R}$ prior to Fourier inversion. Because the far-field patterns $\BE_\infty$ and $\BH_\infty$ recorded on $\Omega_M$ map directly to the spectral components $(f_\bel, g_\bel)$ through Eqs. \eqref{flh}--\eqref{fle0}, restoring missing measurements is equivalent to completing the unobserved Fourier spectrum. Since $\BF$ is geometrically sparse and compactly supported, it can be modeled as a signal with a FRI property \cite{vetterli2002sampling}. Under FRI signal theory, the geometric properties of the source are encapsulated within a sparse set of low-frequency spatial modes. This implies the existence of an annihilating filter in the Fourier domain, which in turn guarantees that the corresponding wrap-around structured Hankel matrix is inherently low-rank \cite{jin2015annihilating, jin2016general}. Consequently, we reframe missing far-field recovery as a constrained low-rank matrix completion problem over a structured Hankel matrix. Once the missing multi-frequency far-field spectrum is enriched at the Nyquist-sampled grid, the Fourier algorithm reconstructs a clean, artifact-free approximation of $\BF$.

\subsection{Constraint Optimization for Data Enrichment}\label{ss:Optimization}

To set up the data enrichment process, we first collect the sparse set of observed spectral coefficients into a single vector. We define the vector representations of the transverse electric ($f_m$) and transverse magnetic ($g_m$) coefficients as
$$
\mathbf{f} = [f_1, f_2, \ldots, f_M]^T, \quad \mathbf{g} = [g_1, g_2, \ldots, g_M]^T,
$$
and concatenate them into a composite spectral vector $\widetilde{\mathbf{s}} := [\mathbf{f}^T, \mathbf{g}^T]^T \in \mathbb{R}^{2M}$. Using $\widetilde{\mathbf{s}}$, we form a block-diagonal Hankel matrix,
\begin{align}
\label{H}
\mathcal{H}_p(\widetilde{\mathbf{s}}) = \begin{bmatrix}  \mathcal{H}_{p_1}(\mathbf{f}) & \mathbf{0}_{M \times p_2} 
\\[0.8em] 
\mathbf{0}_{M \times p_1} & \mathcal{H}_{p_2}(\mathbf{g})  \end{bmatrix} \in \mathbb{R}^{2M \times p},
\end{align}
where $p := p_1 + p_2$ represents the total \textit{matrix pencil size}, and $\mathbf{0}_{M \times p_1}$, and $\mathbf{0}_{M \times p_2}$ are zero matrices. We recall that, for any generic vector $\mathbf{v} = [v_1, v_2, \dots, v_M]^T$ and a chosen matrix pencil size $p$, the corresponding Hankel matrix is defined as,
$$
\mathcal{H}_p(\mathbf{v}) = \begin{bmatrix}  v_1 & v_2 & \dots & v_p \\ v_2 & v_3 & \dots & v_{p+1} \\ \vdots & \vdots & \ddots & \vdots \\ v_M & v_1 & \dots & v_{p-1} \end{bmatrix} \in \mathbb{R}^{M \times p}.
$$

Our core strategy relies on the fact that $\BF$ is spatially compact and geometrically sparse. This physical constraint means its Fourier coefficients satisfy the FRI property. As a result, there exists a non-trivial annihilating filter $\widetilde{\mathbf{h}}$ in the Fourier domain that satisfies the discrete convolution relationship
$$
(\widetilde{\mathbf{h}} * \widetilde{\mathbf{s}})_k := \sum_{i=0}^{n_h} [\widetilde{\mathbf{h}}]_i [\widetilde{\mathbf{s}}]_{k-i} = 0,
$$
where $[\widetilde{\mathbf{h}}]_i$ and $n_h+1$ denote the $i$-th element and the length of the annihilating filter, respectively. The existence of an annihilating filter implies that the wrap-around structured Hankel matrix $\mathcal{H}_p(\widetilde{\mathbf{s}})$ \eqref{H} is rank-deficient. Thanks to the FRI property, selecting an appropriate pencil size $p$ ensures that $\mathcal{H}_p(\widetilde{\mathbf{s}})$ is inherently low-rank. Specifically,  ${\rm rank}(\mathcal{H}_p(\widetilde{\mathbf{s}})=n_h$ if $p$ ($<M$) is chosen larger than the minimum filter size \cite[Theorem II.1]{ye2016compressive}. Exploiting this low-rank structure allows us to cast the missing spectrum recovery problem as a rank-minimization problem subject to data consistency,
\begin{eqnarray}
\begin{array}{ll}
&\arg \ds\min _{\mathbf{s}\in\RR^{2M}} \big(\text{rank}(\mathcal{H}_p(\mathbf{s}))\big)  
\\
&\text{subject to} \quad \mathcal{P}_{\Omega_{M_R}}(\mathbf{s}) = \mathcal{P}_{\Omega_{M_R}}(\widetilde{\mathbf{s}}_{\text{measured}}),\label{e:optimization}
\end{array}
\end{eqnarray}
where $\mathcal{P}_{\Omega_{M_R}}$ is the sampling projection operator that isolates the observed indices in $\Omega_M$. Because direct rank minimization in \eqref{e:optimization} is NP-hard, we use the standard convex relaxation, replacing the rank function with the nuclear norm,
 \begin{eqnarray}
\begin{array}{ll}
 \ds\arg\min_{\mathbf{s}\in\RR^{2M}} \|\mathcal{H}_p(\mathbf{s})\|_*
\\
\text{subject to} \quad \mathcal{P}_{\Omega_{M_R}}(\mathbf{s}) = \mathcal{P}_{\Omega_{M_R}}(\widetilde{\mathbf{s}}_{\text{measured}}).\label{e:optimization2}
\end{array}
\end{eqnarray}
 
 Many algorithms can be used to solve low-rank matrix completion optimization problem and specifically, convex problem \eqref{e:optimization2}. However, we employ ALOHA due to its stability, efficiency, and robustness. ALOHA re-parametrizes  the matrix nuclear norm $\Vert{}\mathcal{H}_p(\mathbf{s})\Vert{}_*$ through a factorized matrix form \cite{jin2015annihilating,jin2016general},
\begin{eqnarray}\label{matrix-factor}
\begin{array}{ll}
\arg\ds\min_{(\mathbf{U},\mathbf{V})\in\mathbb{M}} \frac{1}{2}(\|\mathbf{U}\|_F^2 + \|\mathbf{V}\|_F^2)
\\
\text{subject to} \quad \mathcal{H}_p(\mathbf{\mathbf{\mathbf{s}}}) = \mathbf{UV}^H \quad \text{and} \quad \mathcal{P}_{\Omega_{M_R}}(\mathbf{\mathbf{s}}) = \mathcal{P}_{\Omega_{M_R}}(\mathbf{\mathbf{\widetilde{\mathbf{s}}}}_{\text{measured}}),
\end{array}
\end{eqnarray}
since 
\begin{align*}
    \|\mathcal{H}_p(\mathbf{s})\|_* = \ds\min_{(\mathbf{U},\mathbf{V})\in\mathbb{M}}\left(\|\mathbf{U}\|_F^2+\|\mathbf{V}
\|^2_F\right).
\end{align*}
where $\|\cdot\|_F$ is the Frobenius norm and 
$$
\mathbb{M}=\left\{(\mathbf{U},\mathbf{V})\,|\,\, \mathcal{H}_p(\mathbf{s})=\mathbf{U}\mathbf{V}^H\right\}.
$$
We then solve the factorized model \eqref{matrix-factor} using the ADMM. The factor matrices $\mathbf{U}$ and $\mathbf{V}$ offer a compact low-rank representation of the Hankel matrix, while the ADMM algorithm guarantees stable numerical convergence under non-smooth constraints.

\section{Experimental Setup and Configuration}\label{sec:Setup}

In this section, we provide details of the experimental setup and geometric configurations for evaluation and comparison of the proposed source reconstruction strategy. 

\subsection{Source Models and Nyquist-Sampled Reference}

The reconstruction domain is considered to be the unit cube
$D=\left(-{1}/{2},{1}/{2}\right)^3$. The Fourier truncation order was set to $N=10$. Therefore, each Fourier-coefficient volume has dimensions
$$
    n\times n\times n=21\times21\times21,
    \qquad n:=2N+1, \qquad M=n^3, 
$$
corresponding to $M=n^3=21^3=9261$ Fourier modes. The magnetic far-field data comprise of $9260$ Fourier modes other than the zero-frequency coefficient $f_0$ which is recovered from the mean projection of the exact current onto the polarization direction. The exact source arrays generated by the forward model are sampled on a $50\times50\times50$ Cartesian grid over $D$.

\begin{figure*}[!htb]
    \centering
    \includegraphics[width=0.8\textwidth]{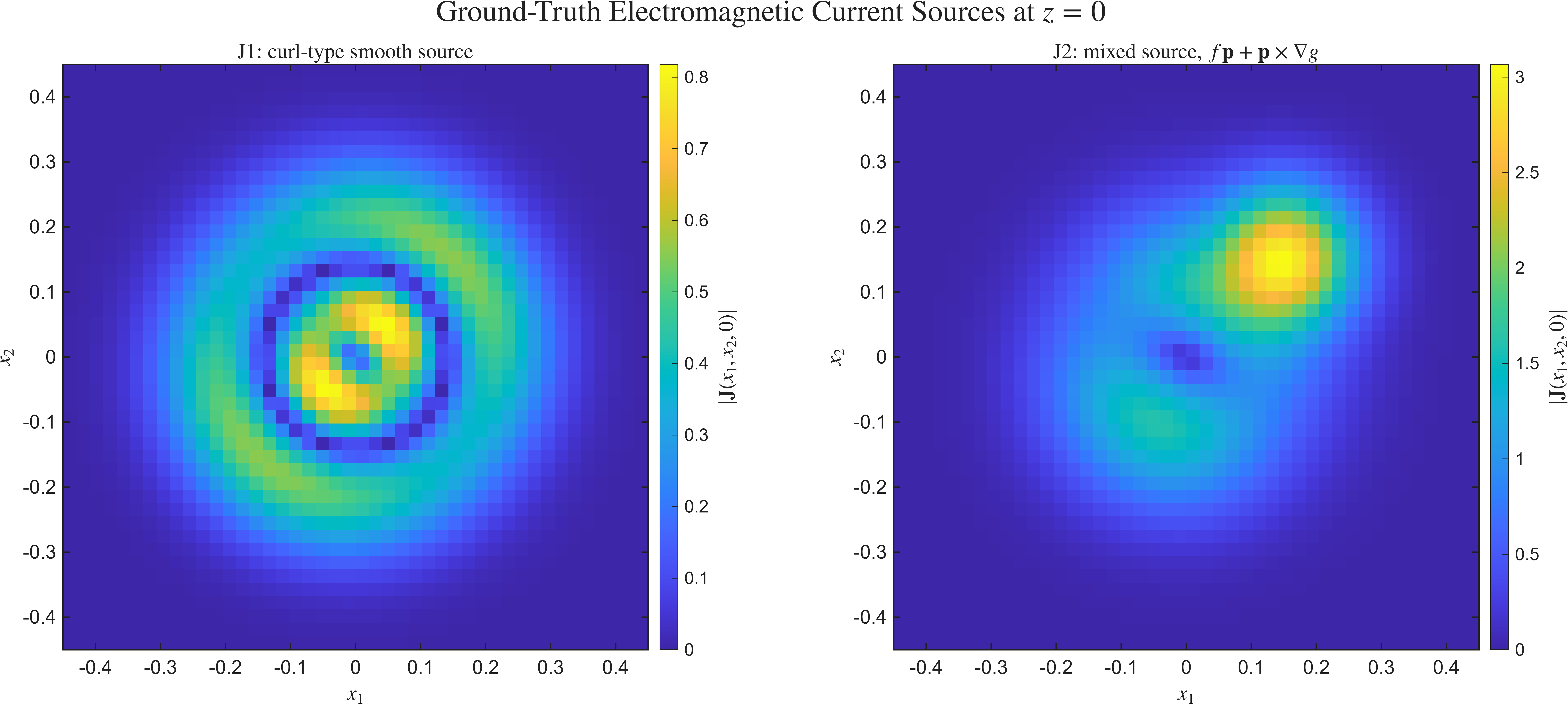}
    \caption{Central-plane representations of electromagnetic-sources.}
    \label{Fig_source_J1_J2_ground_truth}
\end{figure*}

Two source configurations are considered. The first source, denoted by J1, is a smooth curl-type current with
$$
    \BF_{\mathrm{J1}}(\bx)
    =\hat{\Bp}_1\times\nabla g(\bx),
    \qquad
    \hat{\Bp}_1=
    \begin{bmatrix}
    0.5590 & -0.5000 & 0.6614
    \end{bmatrix}^{T},
$$
so that its longitudinal component satisfies $f=0$. The second source, denoted by J2, contains both transverse electric and magnetic components,
$$
    \BF_{\mathrm{J2}}(\bx)
    =\hat{\Bp}_2 f(\bx)+\hat{\Bp}_2\times\nabla g(\bx),
    \qquad
    \hat{\Bp}_2=
    \begin{bmatrix}
    0.7454 & -0.3333 & 0.5774
    \end{bmatrix}^{T}.
$$
J1 therefore tests recovery of a smooth curl-dominated topology, whereas J2 tests a mixed source with spatially separated low- and high-intensity regions. Central-plane representations of electromagnetic-sources are shown in Fig.~\ref{Fig_source_J1_J2_ground_truth}.

The fully sampled Fourier reconstruction is used as the numerical reference for \textit{Peak Signal-to-Noise Ratio} (PSNR) and \textit{Structural Similarity Index} (SSIM). This choice isolates the error introduced by measurement removal and spectral completion from the unavoidable truncation error associated with $N=10$. For visualization, the recovered Fourier series is evaluated on a finer $101\times101\times101$ spatial grid by centered zero-padding of the Fourier coefficient volumes. This operation evaluates the same band-limited Fourier series on a denser grid and does not introduce additional measured frequencies. The central $x_3=0$ slice is used for two-dimensional quantitative evaluation, with the metric field of view restricted to $[-0.4,0.4]^2$.

\subsection{Sparse Sampling and Measurement Noise}

Three measurement rates, $30\%$, $40\%$, and $50\%$, are considered. The percentages are defined relative to the $9260$ nonzero Fourier modes, and the origin is retained in every reconstruction. Random masks are generated in conjugate pairs so that, whenever a mode $\bel$ is selected, its counterpart $-\bel$ is also selected. This preserves the Fourier conjugate symmetry implied by a real-valued spatial current. Within each Monte Carlo trial, the $30\%$, $40\%$, and $50\%$ masks are nested; thus, the lower-rate measurements are strict subsets of the higher-rate measurements. The same mask is used by all reconstruction methods in a given trial and condition.

Two measurement conditions are evaluated:
\begin{enumerate}
    \item noise-free magnetic far-field data; and
    \item magnetic far-field data corrupted by circular complex \textit{Additive White Gaussian Noise} (AWGN) at a \textit{Signal-to-Noise Ratio} (SNR) of $10$ dB.
\end{enumerate}
For the noisy case, the perturbation is applied directly to the complex magnetic far-field array before computing the Fourier coefficients,
$$
    \widetilde{\BH}_{\infty}
    =\BH_{\infty}+\Bn,
    \qquad
    \Bn\sim\mathcal{CN}\left(
    \mathbf{0},
    \frac{P_{\BH}}{10^{\mathrm{SNR}/10}}I
    \right),
$$
where $P_{\BH}$ denotes the mean squared magnitude of the clean far-field data and $I$ is $3\times3$ identity matrix. Independent random seeds are used for the sampling mask and AWGN in every trial. The experiment uses $30$ trials per source, noise condition, and sampling rate. All methods are evaluated on exactly the same realization within a trial, allowing paired comparisons.

\subsection{Compared Reconstruction Methods}
In all numerical simulations, we compare three reconstructions techniques as detailed below.

\subsubsection{Zero-Filled Sub-Sampled Reconstruction.}
The unavailable Fourier coefficients are set to zero, and the inverse Fourier-series synthesis is applied without data completion. This method represents the direct reconstruction from sparse measurements.

\subsubsection{Scale-Normalized Complex $\ell_1$-CS.}
The scalar coefficient volumes $f_{\bel}$ and $g_{\bel}$ are recovered independently using a complex-valued three-dimensional \textit{Discrete Cosine Transform} (DCT) sparsity prior. For a measured coefficient volume $\By$, the baseline solves the \textit{basis-pursuit-denoising} problem
$$
    \min_{\Bx}\ \left\|\mathcal{D}_3\Bx\right\|_1
    \quad\text{subject to}\quad
    \left\|\mathcal{P}_{\Omega_{M_R}}(\Bx-\By)\right\|_2\leq\epsilon,
    %\label{eq:l1_experiment}
$$
where $\mathcal{D}_3$ is an orthonormal separable 3D DCT and $\|\cdot\|_{1}$ and $\|\cdot\|_2$ are the $\ell_1$ and $\ell_2$ norms. The observed coefficients are first normalized by their root-mean-square magnitude, and phase-preserving complex soft thresholding is used in the ADMM iterations. The fixed parameters are $\rho=20$, a maximum of $800$ iterations, and a relative stopping tolerance of $10^{-6}$. For noise-free measurements, $\epsilon=0$. For noisy experiment, with $10$ dB AWGN, the \textit{relative data-fidelity radius} is set to
$$
    \epsilon_{\mathrm{rel}}
    :=\dfrac{1}{\sqrt{10^{\mathrm{SNR}/10}+1}}
    =\dfrac{1}{\sqrt{11}}
    \approx0.3015.
$$

\subsubsection{Joint 3D ALOHA}
The numerical implementation uses a multidimensional extension of the structured-Hankel completion in Section~\ref{ss:Optimization}. Let $\widehat{F}_q\in\mathbb{C}^{n\times n\times n}$ denote the Fourier coefficient volume of the $q$-th current component. The three components are completed jointly through
$$
    \mathcal{H}_{\mathrm{joint}}(\widehat{\BF})
    =
    \left[
    \mathcal{H}_3(\widehat{F}_1),\
    \mathcal{H}_3(\widehat{F}_2),\
    \mathcal{H}_3(\widehat{F}_3)
    \right],
    %\label{eq:joint_hankel_experiment}
$$
where $\mathcal{H}_3$ denotes the 3D block-Hankel lifting. This joint representation exploits the common spatial support of the three electromagnetic current components. A $3\times3\times3$ Hankel patch is used for both sources. The assumed ranks are $22$ for J1 and $30$ for J2; the ADMM penalty is $\mu_0=10$, and $20$ iterations are performed. These parameters are fixed across all measurement rates, noise conditions, and Monte Carlo trials.

Hard measured-data consistency is used for the noise-free case. Under $10$ dB AWGN, the measured-data weight is relaxed to
$$
    \alpha_{\mathrm{data}}
    =\max\left(0.65,1-\epsilon_{\mathrm{rel}}\right)
    \approx0.6985,
$$
which prevents the completion from reproducing every noisy coefficient exactly. After $\ell_1$-CS and ALOHA recovery, conjugate symmetry is enforced before Fourier synthesis.

\subsection{Evaluation Metrics and Statistical Analysis}

For the central slice, PSNR is defined by
$$
    \mathrm{PSNR}
    =
    10\log_{10}
    \left(
    \frac{L^2}{\mathrm{MSE}}
    \right),
    \qquad
    L=\max(\BF_{\mathrm{ref}})-\min(\BF_{\mathrm{ref}}),
$$
where the fully sampled reconstruction is the reference $\BF_{\mathrm{ref}}$ and MSE is the \textit{Mean Squared Error}. SSIM is calculated using an $11\times11$ Gaussian window with standard deviation $1.5$, $C_1=(0.01L)^2$, and $C_2=(0.03L)^2$.

For each method and condition, the reported accuracy is the mean $\pm$ standard deviation over $30$ trials. The error bars in the quantitative plots are two-sided $95\%$ Student-$t$ confidence intervals of the mean. Statistical testing is paired because all methods use the same mask and noise realization in a trial. The primary comparison is Joint 3D ALOHA versus $\ell_1$-CS; Joint 3D ALOHA versus the zero-filled reconstruction is treated as a secondary comparison. At each source, noise level, sampling rate, and metric, a two-sided exact Wilcoxon signed-rank test is applied to the paired differences. Holm correction controls the family-wise error rate at $0.05$ separately within each comparator--metric family. Practical significance is characterized using the mean and median paired improvements, a $10{,}000$-sample percentile-bootstrap $95\%$ confidence interval, paired Cohen's $d_z$, rank-biserial correlation, and the proportion of trials won by ALOHA. Since the sampling masks are nested across rates, statistical tests are conducted separately at each rate and do not assume independence between the $30\%$, $40\%$, and $50\%$ conditions.

Representative full-volume reconstructions are saved for trial~1 at $30\%$ measurements under both noise conditions. The 3D volumes are not averaged across trials because averaging reconstructions generated from different masks would blur source topology. For the 3D figures, all methods within a source use identical physical axes and common isosurface thresholds at $20\%$ and $55\%$ of the fully sampled reference maximum.

\subsection{Runtime Protocol}

Runtime is measured separately from the Monte Carlo experiment to avoid conflating algorithmic time with mask variability and first-call overhead. One warm-up execution is followed by five timed repetitions using the same fixed input. The reported interval includes coefficient recovery and spatial Fourier synthesis but excludes file loading, mask generation, statistical analysis, and figure export. The benchmark was executed under Windows~11 Pro on a 12th-generation Intel Core i7-12700F CPU at $2.10$ GHz with 12 physical cores and $32$ GB of memory.

\section{Numerical Simulations and Results}\label{sec:Num}

In this section, we evaluate the performance of our reconstruction framework across two 3D Maxwell source models subjected to sparse, noise-contaminated multi-frequency measurements. To benchmark our results, we compare the proposed 3D ALOHA-assisted method against two standard baselines: an uncorrected zero-filled Fourier reconstruction and a scale-normalized complex $\ell_1$-CS approach. Rather than relying on a single deterministic test case, all quantitative metrics are calculated over multiple independent Monte Carlo runs and validated through paired statistical analysis.

\subsection{Qualitative Reconstruction at Severe Under-Sampling}\label{ss:qualitative_results}

\begin{figure*}[!htb]
    \centering
    \includegraphics[width=\textwidth]{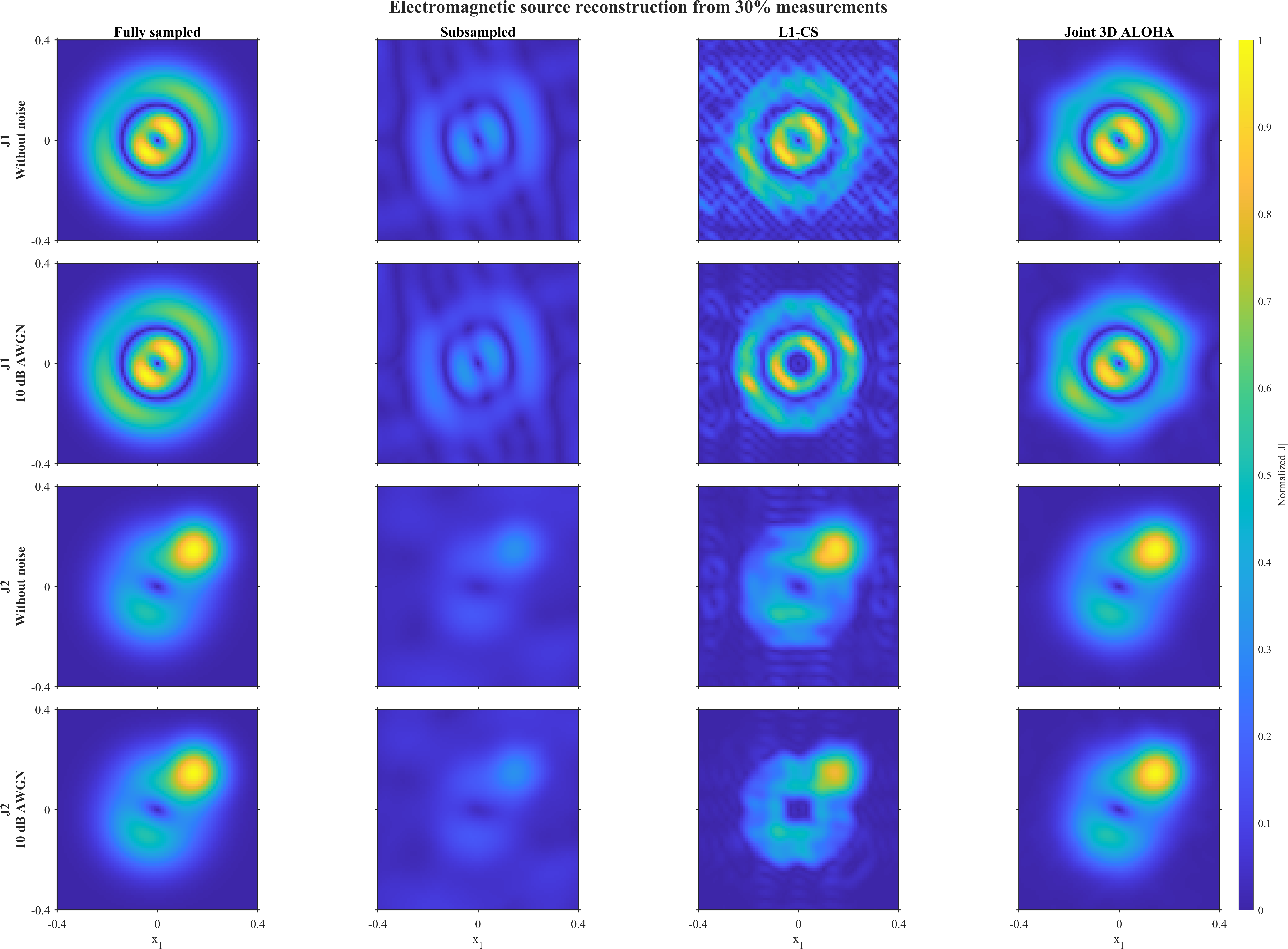}
    \caption{Central-plane electromagnetic-source reconstruction from $30\%$ of the Fourier measurements. Rows from top to bottom: J1 without noise, J1 with $10$ dB AWGN, J2 without noise, and J2 with $10$ dB AWGN. Columns from left to right: fully sampled reference, zero-filled subsampled reconstruction, scale-normalized complex $\ell_1$-CS, and Joint 3D ALOHA. Display intensities are normalized independently by the fully sampled maximum of each row.}
    \label{fig:maxwell_qualitative}
\end{figure*}

Fig.~\ref{fig:maxwell_qualitative} compares the fully sampled reference, zero-filled reconstruction, $\ell_1$-CS, and Joint 3D ALOHA at the most challenging $30\%$ measurement rate. The rows correspond to J1 and J2 under noise-free and $10$ dB AWGN measurements. Each row is normalized by the maximum value of its fully sampled reference for display only; the numerical metrics are calculated from the unnormalized reconstructions.

For J1, direct zero filling suppresses the oscillatory rings and introduces broad low-frequency distortions. The $\ell_1$-CS baseline recovers the dominant central and annular structures but retains structured ripple artifacts, particularly under noise. Joint 3D ALOHA more closely reproduces both the central curl-like region and the outer rings in the noise-free and noisy cases. For J2, zero filling strongly attenuates the two source regions, while $\ell_1$-CS restores their approximate locations but introduces block-like and ringing artifacts. Joint 3D ALOHA preserves the relative shape, separation, and intensity distribution of the two regions, including under $10$ dB AWGN.

\subsection{Quantitative Reconstruction Accuracy}\label{ss:quantitative_results}

\begin{figure*}[!htb]
    \centering
    \includegraphics[width=\textwidth]{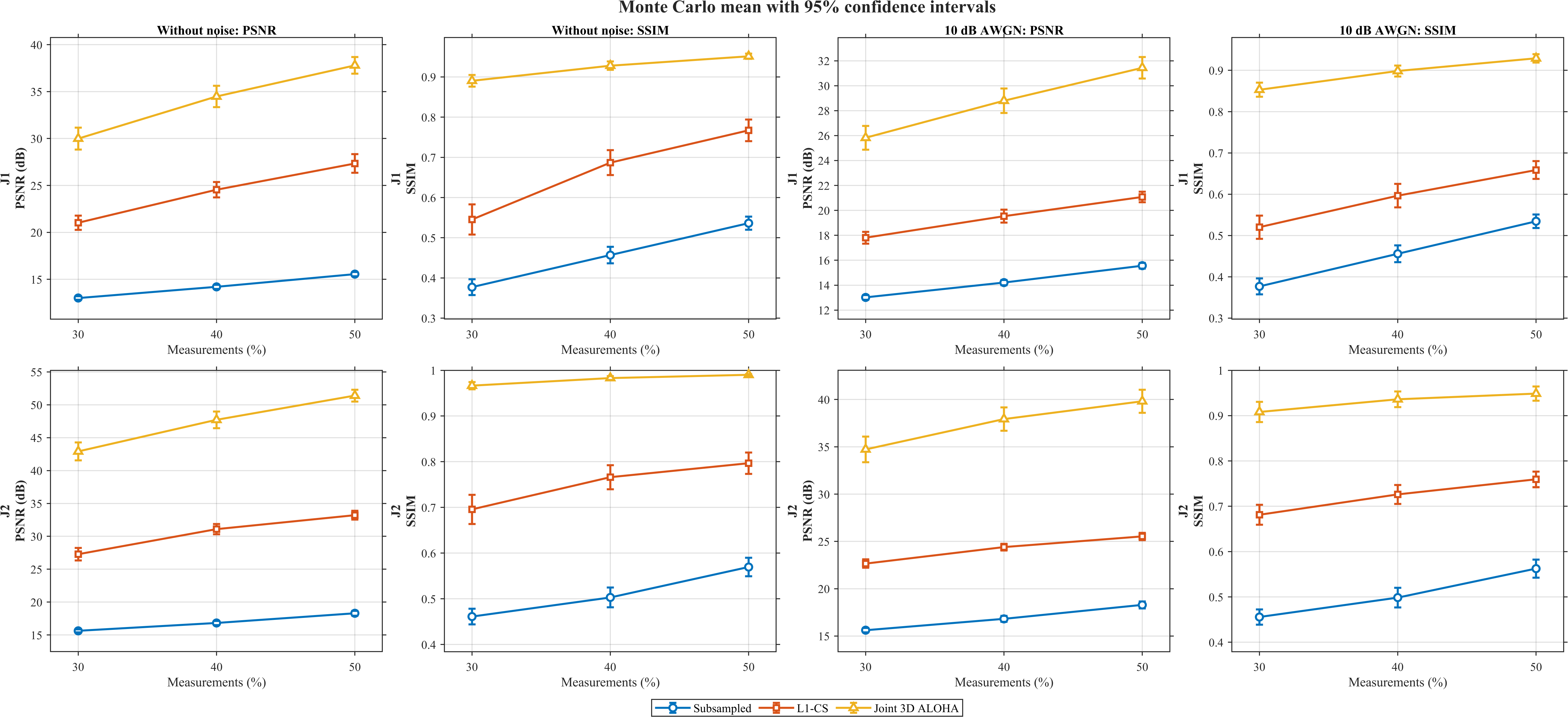}
    \caption{PSNR and SSIM versus measurement rate for J1 (top row) and J2 (bottom row). The first two columns correspond to noise-free data, and the last two columns correspond to $10$ dB AWGN. Markers show the mean over $30$ independent trials, and error bars show two-sided $95\%$ Student-$t$ confidence intervals.}
    \label{fig:maxwell_quantitative}
\end{figure*}

Tables~\ref{tab:maxwell_j1_accuracy} and~\ref{tab:maxwell_j2_accuracy} report the Monte Carlo mean and standard deviation of PSNR and SSIM. Fig.~\ref{fig:maxwell_quantitative} shows the same results as functions of measurement rate, with $95\%$ confidence intervals.
\begin{table*}[!htb]
\centering
\small
\caption{Reconstruction accuracy for J1 (curl-type smooth source) over 30 independent Monte Carlo trials. Values are mean $\pm$ standard deviation. The best result in each row is shown in bold.}
\label{tab:maxwell_j1_accuracy}
\resizebox{\textwidth}{!}{%
\begin{tabular}{llcc|cc|cc}
\toprule
 & & \multicolumn{2}{c|}{Sub-sampled} & \multicolumn{2}{c|}{$\ell_1$-CS} & \multicolumn{2}{c}{Joint 3D ALOHA} \\
Scenario & Measurements (\%) & PSNR (dB) & SSIM & PSNR (dB) & SSIM & PSNR (dB) & SSIM \\
\midrule
Without noise & 30 & $13.01 \pm 0.51$ & $0.3772 \pm 0.0528$ & $21.02 \pm 2.02$ & $0.5454 \pm 0.1007$ & $\mathbf{30.00 \pm 3.13}$ & $\mathbf{0.8905 \pm 0.0388}$ \\
 & 40 & $14.19 \pm 0.58$ & $0.4569 \pm 0.0551$ & $24.55 \pm 2.18$ & $0.6869 \pm 0.0831$ & $\mathbf{34.48 \pm 3.06}$ & $\mathbf{0.9280 \pm 0.0281}$ \\
 & 50 & $15.55 \pm 0.61$ & $0.5362 \pm 0.0439$ & $27.34 \pm 2.66$ & $0.7671 \pm 0.0723$ & $\mathbf{37.79 \pm 2.39}$ & $\mathbf{0.9513 \pm 0.0174}$ \\
\midrule
10 dB AWGN & 30 & $13.02 \pm 0.51$ & $0.3768 \pm 0.0520$ & $17.80 \pm 1.26$ & $0.5201 \pm 0.0756$ & $\mathbf{25.82 \pm 2.55}$ & $\mathbf{0.8531 \pm 0.0456}$ \\
 & 40 & $14.21 \pm 0.58$ & $0.4558 \pm 0.0547$ & $19.54 \pm 1.39$ & $0.5966 \pm 0.0762$ & $\mathbf{28.80 \pm 2.63}$ & $\mathbf{0.8984 \pm 0.0357}$ \\
 & 50 & $15.56 \pm 0.61$ & $0.5345 \pm 0.0436$ & $21.08 \pm 1.13$ & $0.6586 \pm 0.0579$ & $\mathbf{31.44 \pm 2.32}$ & $\mathbf{0.9291 \pm 0.0274}$\\
\bottomrule
\end{tabular}}
\end{table*}

\begin{table*}[!htb]
\centering
\small
\caption{Reconstruction accuracy for J2 (mixed $\hat{\Bp}f+\hat{\Bp}\times\nabla g$ source) over 30 independent Monte Carlo trials. Values are mean $\pm$ standard deviation. The best result in each row is shown in bold.}
\label{tab:maxwell_j2_accuracy}
\resizebox{\textwidth}{!}{%
\begin{tabular}{llcc|cc|cc}
\toprule
 & & \multicolumn{2}{c|}{Subsampled} & \multicolumn{2}{c|}{$\ell_1$-CS} & \multicolumn{2}{c}{Joint 3D ALOHA} \\
Scenario & Measurements (\%) & PSNR (dB) & SSIM & PSNR (dB) & SSIM & PSNR (dB) & SSIM \\
\midrule
Without noise & 30 & $15.62 \pm 0.62$ & $0.4611 \pm 0.0457$ & $27.29 \pm 2.55$ & $0.6957 \pm 0.0856$ & $\mathbf{42.91 \pm 3.65}$ & $\mathbf{0.9664 \pm 0.0213}$ \\
 & 40 & $16.82 \pm 0.85$ & $0.5030 \pm 0.0581$ & $31.09 \pm 2.07$ & $0.7661 \pm 0.0703$ & $\mathbf{47.71 \pm 3.40}$ & $\mathbf{0.9832 \pm 0.0112}$ \\
 & 50 & $18.30 \pm 0.99$ & $0.5695 \pm 0.0541$ & $33.22 \pm 1.78$ & $0.7968 \pm 0.0626$ & $\mathbf{51.40 \pm 2.40}$ & $\mathbf{0.9901 \pm 0.0049}$\\
\midrule
10 dB AWGN & 30 & $15.62 \pm 0.62$ & $0.4557 \pm 0.0447$ & $22.66 \pm 1.19$ & $0.6812 \pm 0.0589$ & $\mathbf{34.73 \pm 3.62}$ & $\mathbf{0.9082 \pm 0.0596}$ \\
 & 40 & $16.81 \pm 0.84$ & $0.4986 \pm 0.0582$ & $24.40 \pm 0.95$ & $0.7261 \pm 0.0562$ & $\mathbf{37.93 \pm 3.32}$
 & $\mathbf{0.9361 \pm 0.0465}$\\
 & 50 & $18.29 \pm 0.98$ & $0.5624 \pm 0.0534$ & $25.53 \pm 1.02$ & $0.7594 \pm 0.0458$ & $\mathbf{39.81 \pm 3.26}$ & $\mathbf{0.9486 \pm 0.0420}$ \\
\bottomrule
\end{tabular}}
\end{table*}

For J1 without noise, increasing the measurement rate from $30\%$ to $50\%$ raises the Joint 3D ALOHA PSNR from $30.00$ to $37.79$ dB and SSIM from $0.8905$ to $0.9513$. Over the same range, $\ell_1$-CS increases PSNR from $21.02$ to $27.34$ dB and SSIM from $0.5454$ to $0.7671$. At $10$ dB AWGN, ALOHA maintains $25.82$--$31.44$ dB PSNR and $0.8531$--$0.9291$ SSIM, compared with $17.80$--$21.08$ dB PSNR and $0.5201$--$0.6586$ SSIM for $\ell_1$-CS.

The separation is larger for J2. In the noise-free experiment, ALOHA achieves $42.91$, $47.71$, and $51.40$ dB PSNR at $30\%$, $40\%$, and $50\%$ measurements, respectively, whereas $\ell_1$-CS achieves $27.29$, $31.09$, and $33.22$ dB PSNR, respectively. The corresponding ALOHA SSIM values are $0.9664$, $0.9832$, and $0.9901$. Under $10$ dB AWGN, the ALOHA PSNR remains between $34.73$ and $39.81$ dB, with SSIM between $0.9082$ and $0.9486$. These results show that the benefit is not limited to the noise-free setting and remains substantial when the far-field data are corrupted before Fourier-coefficient computation.

All methods improve as the sampling rate increases. However, the zero-filled reconstruction improves only gradually without recovering the missing coefficients. The $\ell_1$-CS baseline benefits more strongly from additional measurements but remains sensitive to the mismatch between DCT sparsity and the structured Maxwell coefficient volumes. Joint ALOHA consistently provides the highest PSNR and SSIM because it uses both structured Hankel low-rank and the shared support of the vector-current components.

\subsection{Paired Significance and Effect-Size Analysis}\label{ss:significance_results}

\begin{figure*}[!htb]
    \centering
    \includegraphics[width=\textwidth]{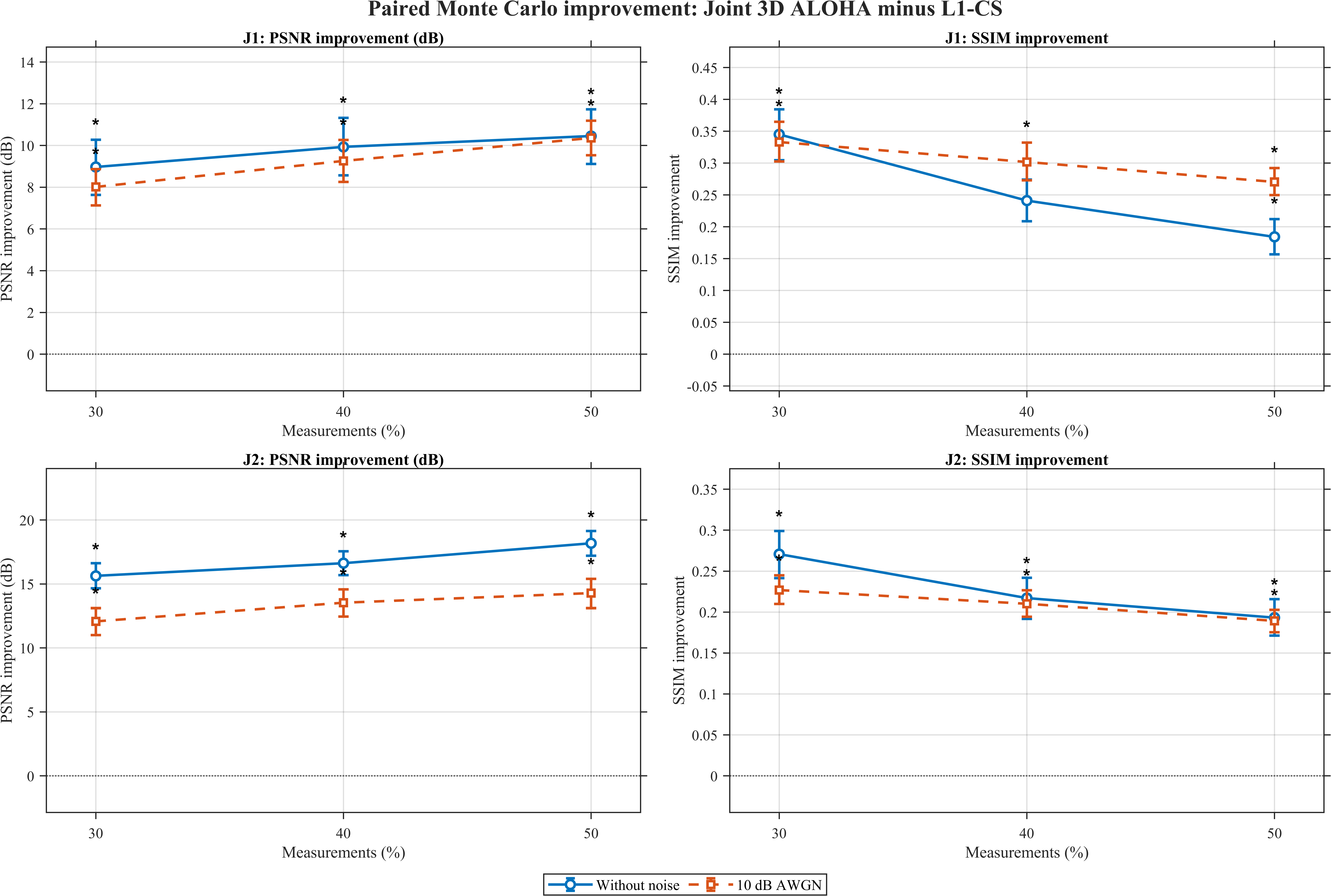}
    \caption{Paired Monte Carlo improvement of Joint 3D ALOHA over $\ell_1$-CS. Error bars show percentile-bootstrap $95\%$ confidence intervals of the mean paired improvement. Asterisks denote significance after Holm correction.}
    \label{fig:maxwell_paired_improvement}
\end{figure*}

Fig.~\ref{fig:maxwell_paired_improvement} visualizes the paired improvement of Joint 3D ALOHA over $\ell_1$-CS. Table~\ref{tab:maxwell_significance} gives the corresponding bootstrap confidence intervals, paired effect sizes, and Holm-adjusted exact Wilcoxon probabilities.
\begin{table*}[!htb]
\centering
\scriptsize
\caption{Paired comparison of Joint 3D ALOHA against $\ell_1$-CS. $\Delta$ denotes ALOHA minus $\ell_1$-CS. Brackets contain percentile-bootstrap 95\% confidence intervals for the paired mean improvement. $d_z$ is paired Cohen's effect size and $p_{\mathrm H}$ is the Holm-adjusted exact two-sided Wilcoxon signed-rank $p$-value.}
\label{tab:maxwell_significance}
\resizebox{\textwidth}{!}{%
\begin{tabular}{lllccc|ccc}
\toprule
 & & & \multicolumn{3}{c|}{PSNR} & \multicolumn{3}{c}{SSIM} \\
Source & Scenario & Meas. (\%) & $\Delta$ (dB) [95\% CI] & $d_z$ & $p_{\mathrm H}$ & $\Delta$ [95\% CI] & $d_z$ & $p_{\mathrm H}$ \\
\midrule
J1 & No noise & 30 & $+8.97$ [+7.63, +10.27] & $2.38$ & $2.24\times10^{-8}$ & $+0.3451$ [+0.3044, +0.3843] & $3.04$ & $2.24\times10^{-8}$ \\
 &  & 40 & $+9.93$ [+8.57, +11.33] & $2.50$ & $2.24\times10^{-8}$ & $+0.2411$ [+0.2087, +0.2740] & $2.60$ & $2.24\times10^{-8}$ \\
 &  & 50 & $+10.45$ [+9.12, +11.73] & $2.85$ & $2.24\times10^{-8}$ & $+0.1842$ [+0.1568, +0.2122] & $2.33$ & $2.24\times10^{-8}$ \\
\addlinespace[2pt]
 & 10 dB AWGN & 30 & $+8.02$ [+7.13, +8.87] & $3.20$ & $2.24\times10^{-8}$ & $+0.3330$ [+0.3026, +0.3648] & $3.79$ & $2.24\times10^{-8}$ \\
 &  & 40 & $+9.26$ [+8.25, +10.26] & $3.22$ & $2.24\times10^{-8}$ & $+0.3017$ [+0.2729, +0.3323] & $3.54$ & $2.24\times10^{-8}$ \\
 &  & 50 & $+10.36$ [+9.53, +11.18] & $4.35$ & $2.24\times10^{-8}$ & $+0.2704$ [+0.2497, +0.2922] & $4.41$ & $2.24\times10^{-8}$ \\
\midrule
J2 & No noise & 30 & $+15.63$ [+14.66, +16.62] & $5.66$ & $2.24\times10^{-8}$ & $+0.2708$ [+0.2413, +0.2989] & $3.30$ & $2.24\times10^{-8}$ \\
 &  & 40 & $+16.62$ [+15.69, +17.55] & $6.25$ & $2.24\times10^{-8}$ & $+0.2171$ [+0.1916, +0.2418] & $3.02$ & $2.24\times10^{-8}$ \\
 &  & 50 & $+18.18$ [+17.21, +19.13] & $6.67$ & $2.24\times10^{-8}$ & $+0.1932$ [+0.1713, +0.2157] & $3.02$ & $2.24\times10^{-8}$ \\
\addlinespace[2pt]
 & 10 dB AWGN & 30 & $+12.07$ [+11.00, +13.11] & $3.97$ & $2.24\times10^{-8}$ & $+0.2270$ [+0.2099, +0.2448] & $4.66$ & $2.24\times10^{-8}$ \\
 &  & 40 & $+13.53$ [+12.45, +14.57] & $4.51$ & $2.24\times10^{-8}$ & $+0.2101$ [+0.1942, +0.2267] & $4.60$ & $2.24\times10^{-8}$ \\
 &  & 50 & $+14.29$ [+13.11, +15.40] & $4.39$ & $2.24\times10^{-8}$ & $+0.1892$ [+0.1755, +0.2027] & $4.98$ & $2.24\times10^{-8}$ \\
\bottomrule
\end{tabular}}
\end{table*}

Across all 12 source--noise--sampling conditions, the paired PSNR improvement over $\ell_1$-CS ranges from $8.02$ to $18.18$ dB, and the paired SSIM improvement ranges from $0.1842$ to $0.3451$. None of the bootstrap intervals includes zero. The paired Cohen's $d_z$ ranges from $2.38$ to $6.67$ for PSNR and from $2.33$ to $4.98$ for SSIM, indicating that the performance separation is large relative to the trial-to-trial variability. ALOHA outperforms $\ell_1$-CS in all 30 paired trials for every tested condition, giving a win rate and rank-biserial correlation of $1.0$. Consequently, every primary comparison has an exact two-sided Wilcoxon $p$-value of $1.86\times10^{-9}$ and a Holm-adjusted value of $2.24\times10^{-8}$. Thus, the observed gains are both statistically reliable and practically substantial.

The PSNR advantage generally increases with measurement rate, especially for J2, because the additional observed coefficients allow the low-rank model to reconstruct finer source detail. The SSIM advantage decreases moderately with rate in the noise-free cases because $\ell_1$-CS also recovers more of the gross structure as additional measurements become available. Nevertheless, the SSIM confidence intervals remain strictly positive at every rate.

\subsection{Full 3D Reconstruction}\label{ss:three_dimensional_results}

\begin{figure*}[!htb]
    \centering
    \includegraphics[width=\textwidth]{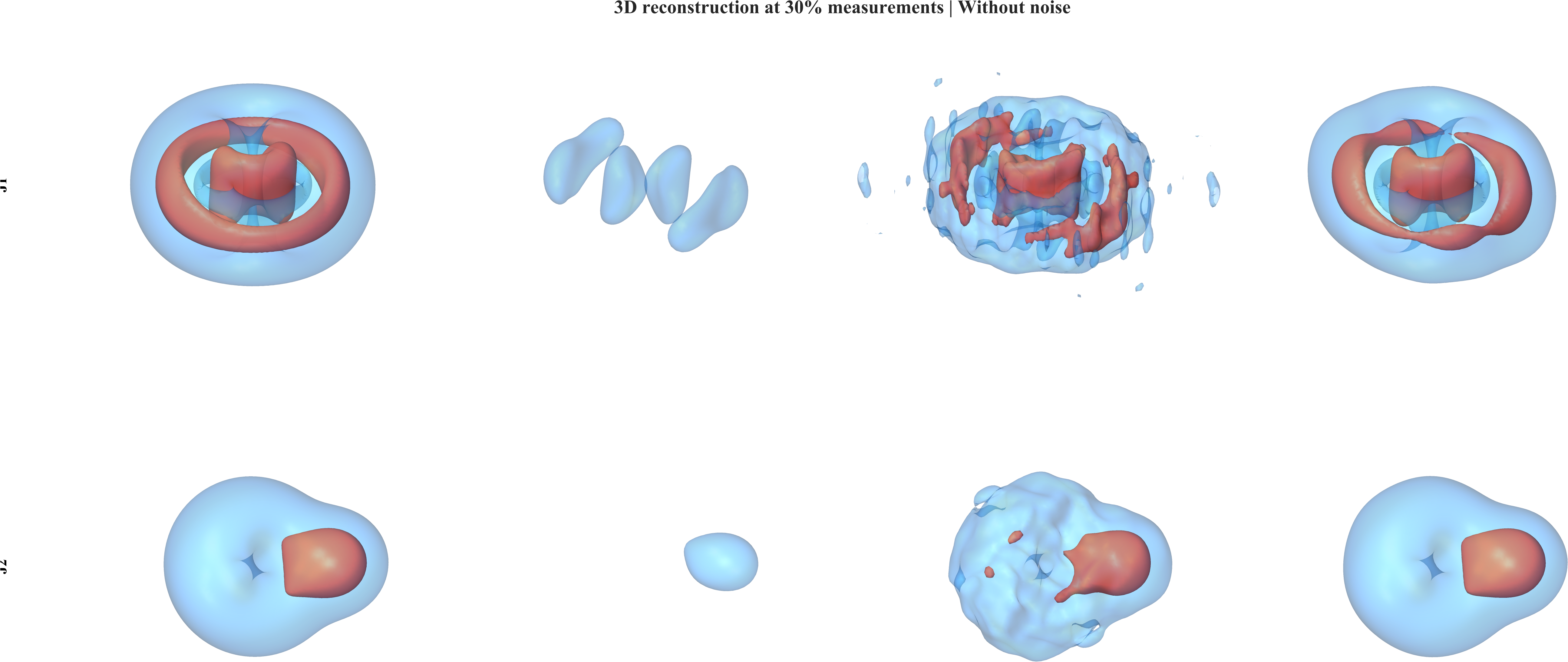}
    \caption{Representative full-volume reconstruction at $30\%$ measurements without noise. Rows correspond to J1 and J2; columns correspond to the fully sampled reference, zero-filled subsampled reconstruction, $\ell_1$-CS, and Joint 3D ALOHA. The blue and red isosurfaces are drawn at $20\%$ and $55\%$ of the fully sampled maximum, respectively. Ground truth in the figure denotes the fully sampled Fourier reference.}
    \label{fig:maxwell_3d_clean}
\end{figure*}

\begin{figure*}[!htb]
    \centering
    \includegraphics[width=\textwidth]{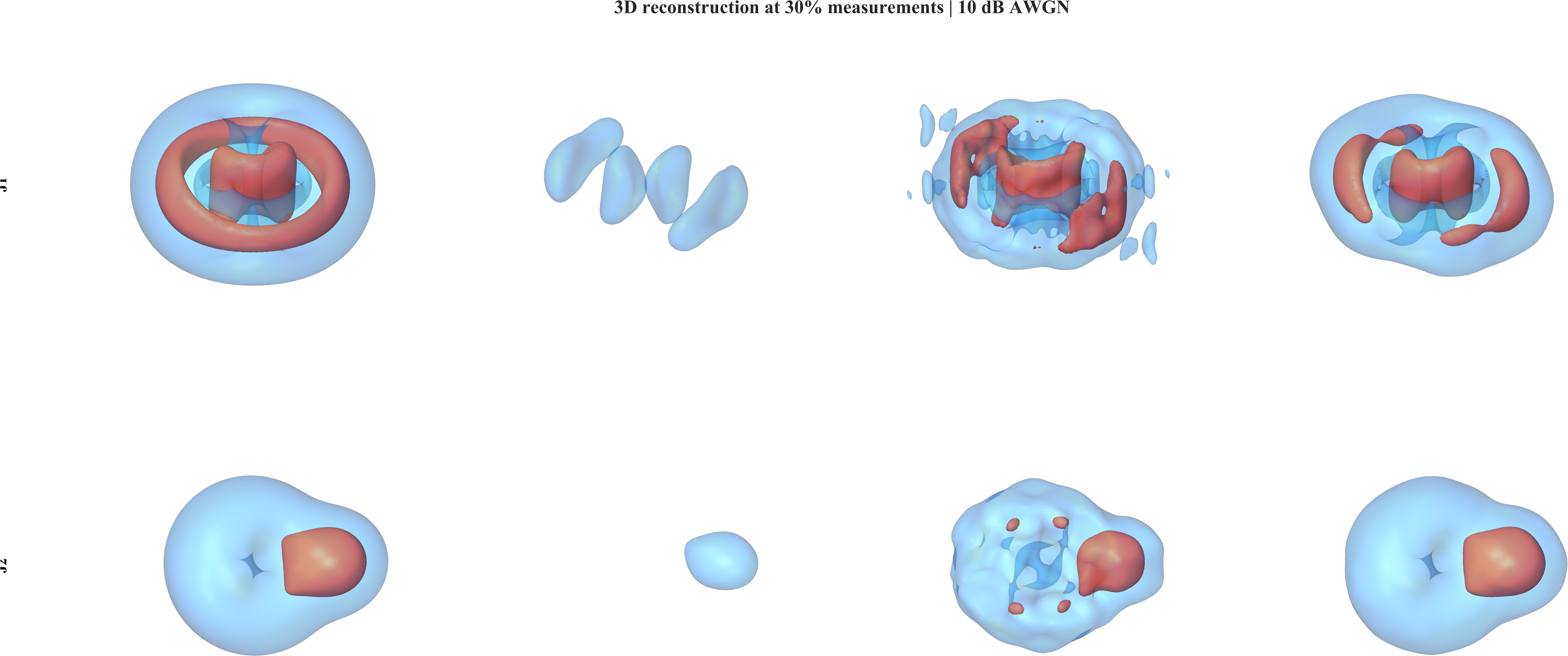}
    \caption{Representative full-volume reconstruction at $30\%$ measurements with $10$ dB AWGN. The row, column, axis, and isosurface conventions are identical to Fig.~\ref{fig:maxwell_3d_clean}.}
    \label{fig:maxwell_3d_noisy}
\end{figure*}

Figs.~\ref{fig:maxwell_3d_clean} and~\ref{fig:maxwell_3d_noisy} compare the full-volume reconstructions at $30\%$ measurements. Within each source, the fully sampled, zero-filled, $\ell_1$-CS, and ALOHA panels use the same physical field of view and the same absolute isosurface thresholds. Therefore, missing or attenuated structures cannot be hidden by independently rescaling each method.

For J1, the zero-filled result breaks the connected toroidal and inner structures into several disconnected lobes. The $\ell_1$-CS result restores more of the volume but contains numerous spurious shells and fragments. ALOHA recovers the principal outer and inner topology with substantially fewer false structures. For J2, direct sub-sampling retains only a small high-intensity region, and $\ell_1$-CS reconstructs the principal components together with irregular low-intensity surfaces. The ALOHA volume closely follows the fully sampled outer support and high-intensity component. The same qualitative behavior remains visible after adding $10$ dB AWGN.

For the representative trial used in the 3D figures, the central-slice PSNR/SSIM values of ALOHA are $29.01$ dB/$0.8917$ for clean J1, $25.74$ dB/$0.8753$ for noisy J1, $46.76$ dB/$0.9758$ for clean J2, and $38.80$ dB/$0.9681$ for noisy J2. These values are representative visual examples rather than Monte Carlo averages; the statistical conclusions are based on the complete 30-trial results in Tables~\ref{tab:maxwell_j1_accuracy}--\ref{tab:maxwell_significance}.

\subsection{Computational Cost}\label{ss:runtime_results}

The warm-up-controlled timing results are provided in Table~\ref{tab:maxwell_runtime}. The direct zero-filled reconstruction requires approximately $0.056$--$0.059$ s because it performs only Fourier synthesis. Joint 3D ALOHA requires $0.666$--$0.720$ s across the tested sources and conditions, whereas the complex DCT $\ell_1$-CS baseline requires $1.236$--$1.737$ s. Thus, in this MATLAB implementation, ALOHA uses approximately $40\%$--$58\%$ of the $\ell_1$-CS runtime, corresponding to a speedup of approximately $1.74$--$2.51$ relative to $\ell_1$-CS. ALOHA remains slower than direct zero filling, but the additional cost produces a large gain in reconstruction accuracy and source topology.

\begin{table*}[!htb]
\centering
\scriptsize
\caption{Warm-up-controlled wall-clock runtime. Each value is mean $\pm$ standard deviation over five executions of the same fixed input after one warm-up execution. Data loading, mask generation, and figure export are excluded.}
\label{tab:maxwell_runtime}
\resizebox{\textwidth}{!}{%
\begin{tabular}{lllccc}
\toprule
Source & Scenario & Meas. (\%) & Subsampled (s) & $\ell_1$-CS (s) & Joint 3D ALOHA (s) \\
\midrule
J1 & Without noise & 30 & $0.057 \pm 0.001$ & $1.345 \pm 0.004$ & $0.666 \pm 0.002$ \\
 &  & 40 & $0.056 \pm 0.001$ & $1.423 \pm 0.021$ & $0.684 \pm 0.010$ \\
 &  & 50 & $0.058 \pm 0.001$ & $1.737 \pm 0.013$ & $0.692 \pm 0.002$ \\
\addlinespace[2pt]
 & 10 dB AWGN & 30 & $0.059 \pm 0.000$ & $1.652 \pm 0.020$ & $0.690 \pm 0.008$ \\
 &  & 40 & $0.058 \pm 0.001$ & $1.693 \pm 0.007$ & $0.689 \pm 0.004$ \\
 &  & 50 & $0.059 \pm 0.001$ & $1.682 \pm 0.012$ & $0.692 \pm 0.003$ \\
\midrule
J2 & Without noise & 30 & $0.058 \pm 0.002$ & $1.558 \pm 0.011$ & $0.717 \pm 0.003$ \\
 &  & 40 & $0.058 \pm 0.000$ & $1.651 \pm 0.013$ & $0.717 \pm 0.002$ \\
 &  & 50 & $0.058 \pm 0.001$ & $1.731 \pm 0.024$ & $0.720 \pm 0.003$ \\
\addlinespace[2pt]
 & 10 dB AWGN & 30 & $0.059 \pm 0.001$ & $1.609 \pm 0.019$ & $0.713 \pm 0.004$ \\
 &  & 40 & $0.059 \pm 0.001$ & $1.236 \pm 0.009$ & $0.711 \pm 0.005$ \\
 &  & 50 & $0.059 \pm 0.001$ & $1.270 \pm 0.015$ & $0.716 \pm 0.002$ \\
\bottomrule
\end{tabular}}
\end{table*}

The runtime comparison is implementation- and hardware-dependent. In particular, the $\ell_1$-CS solver uses up to $800$ ADMM iterations separately for $f$ and $g$, while ALOHA uses $20$ joint factorization iterations. Consequently, the timing result should be interpreted as a comparison of the tested implementations and fixed parameters rather than a universal complexity ordering.

\subsection{Discussion and Limitations}\label{ss:discussion_results}

The Monte Carlo experiments establish three main observations. First, joint structured-Hankel completion provides a large advantage at severe under-sampling, with the strongest absolute PSNR gains observed for the mixed J2 source. Second, the advantage persists under $10$ dB complex AWGN and is not attributable to a favorable single mask or noise realization. Third, the improvement extends from central-slice metrics to preservation of the full three-dimensional source topology.

The comparison also has limitations. Only two simulated source families are evaluated, and both use the same Fourier--Maxwell forward model employed to construct the reconstruction data. The noise study is limited to AWGN at one SNR and does not include calibration errors, polarization mismatch, nonuniform frequency loss, or modeling error. The $\ell_1$-CS baseline uses a complex DCT sparsity model; other regularizers, such as joint spatial group sparsity or total variation, may provide stronger alternatives for some source classes. Finally, the ALOHA rank is source-specific, although it is fixed before testing and is not changed across sampling rates, noise conditions, or Monte Carlo trials. Future work should examine automatic rank selection, broader source geometries, experimental far-field measurements, and robustness to forward-model mismatch.

\section{Conclusions}\label{sec:Conc}

A two-stage data enrichment strategy has been investigated for reconstructing three-dimensional electromagnetic current sources from sparse multi-frequency magnetic far-field data. Missing Fourier coefficients are first estimated using joint three-dimensional structured-Hankel completion, after which the current is synthesized using the Fourier reconstruction formula. The joint ALOHA realization exploits the common support of the three vector-current components and is compared with direct zero filling and a scale-normalized complex $\ell_1$-CS baseline.

The numerical study uses two Maxwell source models, $30\%$--$50\%$ conjugate-paired measurements, noise-free and $10$ dB AWGN conditions, and $30$ independent Monte Carlo trials. Joint 3D ALOHA achieves paired PSNR gains of $8.02$--$18.18$ dB and SSIM gains of $0.1842$--$0.3451$ over $\ell_1$-CS. All paired gains remain significant after Holm correction ($p_{\mathrm H}=2.24\times10^{-8}$), with large paired effect sizes and a $100\%$ trial win rate. The full-volume reconstructions further show that the method preserves the principal three-dimensional source topology more faithfully than zero filling or the tested $\ell_1$-CS baseline. In the warm-up-controlled MATLAB benchmark, Joint 3D ALOHA requires approximately $0.67$--$0.72$ s per reconstruction and is $1.74$--$2.51$ times faster than the implemented complex DCT $\ell_1$-CS solver.

These results support structured-Hankel data enrichment as an effective approach for sparse electromagnetic inverse source reconstruction. Further validation on additional source geometries, measurement-noise models, forward-model mismatch, and experimental far-field data is required before drawing conclusions about broader deployment settings.

%\section*{Competing Interests}

%The author(s) declared no potential conflicts of interest with respect to the research, authorship, and/or publication of this article.

%\section*{Declaration of Generative AI and AI-assisted Technologies in the Manuscript Preparation Process.}
%During the preparation of this work the authors used Gemini.ai (Google) in order to improve the readability and language of their own writing. After using this tool, the authors reviewed and edited the content as needed and take full responsibility for the content of the published article.
	
\bibliographystyle{IEEEtran}

\end{document}